\documentclass[12pt]{article}
\pdfoutput=1

\usepackage[bulletsep]{collref}
\usepackage{amssymb,graphicx}
\usepackage[intlimits]{amsmath}
\usepackage{bbm}
\usepackage[small]{subfigure}

\usepackage{MnSymbol}

\makeatletter \@addtoreset{equation}{section} \makeatother

\makeatletter
\let\old@startsection=\@startsection
\let\oldl@section=\l@section
\renewcommand{\@startsection}[6]{\old@startsection{#1}{#2}{#3}{#4}{#5}{#6\mathversion{bold}}}
\renewcommand{\l@section}[2]{\oldl@section{\mathversion{bold}#1}{#2}}
\makeatother

\makeatletter
\let\old@makecaption=\@makecaption
\def\@makecaption{\small\old@makecaption}
\makeatother

\begin{document}


\thispagestyle{empty}
\today
\begin{flushright}\footnotesize
\texttt{}\\
\texttt{} \\
\texttt{}

\vspace{0.3cm}
\end{flushright}

\renewcommand{\thefootnote}{\fnsymbol{footnote}}
\setcounter{footnote}{0}

\begin{center}
{\Large\textbf{\mathversion{bold} 
Matching quantum string corrections and circular Wilson loops in $AdS_4 \times CP^3$     
}
\par}

\vspace{0.3cm}

\textrm{Daniel~Medina-Rincon}
\vspace{4mm}

{\small 
\textit{Institut f\"ur Theoretische Physik,\\
Eidgen\"ossische Technische Hochschule Z\"urich,\\
Wolfgang-Pauli-Strasse 27, 8093 Z\"urich, Switzerland}\\

\vspace{0.2cm}
\texttt{medina.daniel@itp.phys.ethz.ch}
}



\par\vspace{0.5cm}

\textbf{Abstract} \vspace{3mm}

\begin{minipage}{13cm}
Recent progresses in the computation of quantum string corrections to holographic Wilson loops are extended to the case of strings in $AdS_{4}\times CP^{3}$. For this, the ratio of $\frac{1}{2}$-BPS circular and $\frac{1}{6}$-BPS latitude fermionic Wilson loops in ABJM is considered at strong coupling by studying the corresponding semiclassical string partition functions. Explicit evaluation of fluctuation determinants using phaseshifts and diffeomorphism invariant regulators leads to exact matching with the recent field theory proposal. Key to this computation is the choice of boundary conditions for massless fermions. In the limit for which the latitude Wilson loop has a trivial expectation value, the long known localization result for the $\frac{1}{2}$-BPS fermionic circular Wilson loop in ABJM is recovered.
\end{minipage}

\end{center}

\vspace{1.5cm}


\def \la{\label}
 \def \ha {\te {1\ov 2}}
\def \te {\textstyle}\def \bi {\bibitem} 
\def \ve {\Lambda}
\def \s {\sigma} \def \ov {\over} 
\def \ci {\cite} \def \td {\tilde} 
  \def \a {\alpha}
\def \iffa {\iffalse} \def \N {{\cal N}} 
\def \foot {\footnote}
\def \ed {\end{document}}
\def \be {\begin{equation}} \def \OO {{\cal O}} 
\def \ee {\end{equation}}  \def \CC  {{\rm C}} \def \sql {\sqrt \lambda} \def \nn {{n}}\def \LL {{\rm L}} \def \l  {\lambda}
\setcounter{page}{1}
\renewcommand{\thefootnote}{\arabic{footnote}}
\setcounter{footnote}{0}

\newpage



\section{Introduction}

The gauge-string correspondence has been one of the most exciting developments in theoretical physics as it allows for insights into the strongly coupled dynamics of gauge theories by considering the corresponding weakly coupled strings. The best known examples of the correspondence are the dualities between $\mathcal{N}=4$ super Yang-Mills with $SU(N)$ gauge group and type IIB strings in $AdS_{5}\times S^{5}$ \cite{Maldacena:1997re}, and between $\mathcal{N}=6$ Chern-Simons matter theory with gauge group $U(N)_{k}\times U(N)_{-k}$ and type IIA string theory in $AdS_{4}\times CP^{3}$ \cite{Aharony:2008ug}.\\
Wilson loops have played an important role within the gauge-string correspondence as they have a precise description at both sides of the duality \cite{Maldacena:1998im,Rey:1998bq}. Localization \cite{Pestun:2007rz} has lead to remarkable progress in the field theory computation of Wilson loops, leading to exact results for all orders in the coupling for certain configurations in supersymmetric field theories. On the string theory side, the large $N$ Wilson loop expectation value corresponds to the superstring partition function \cite{Polyakov:1997tj} and only the leading contribution, and in only few cases the next to leading order contribution, has been matched with field theory at strong coupling. 

For the case of $\mathcal{N}=4$ SYM there exist exact results at all orders in the coupling for the circular \cite{Pestun:2007rz,Erickson:2000af,Drukker:2000rr} and latitude Wilson loops \cite{Drukker:2006ga}. From the string theory side in $AdS_{5}\times S^{5}$ at 1-loop, based on the foundational work in \cite{Drukker:2000ep}, early attempts at matching the field theory result in the planar limit for the circular Wilson loop lead to discrepancies attributed to the normalization of the string path integral measure \cite{Kruczenski:2008zk,Kristjansen:2012nz,Buchbinder:2014nia}. Renewed interest in this program lead to the proposal of the 1-loop string theory computation of the ratio of $\frac{1}{2}$-BPS circular and $\frac{1}{4}$-BPS latitude Wilson loops, in which measure factors would play no role \cite{Forini:2015bgo,Faraggi:2016ekd}. Despite early discrepancies  \cite{Forini:2015bgo,Faraggi:2016ekd,Aguilera-Damia:2018twq}, a precise matching of this ratio was achieved at the first two leadings orders in strong coupling: Initially only at first order in small values of the latitude parameter \cite{Forini:2017whz}, and finally at all orders in the latitude parameter \cite{Cagnazzo:2017sny}. Recently, the $\frac{1}{2}$-BPS circular Wilson loop was successfully computed at 1-loop in string theory at strong coupling by considering the ratio between the semiclassical string partition function dual to this configuration and the one dual to a latitude Wilson loop with trivial expectation value \cite{Medina-Rincon:2018wjs}.

In $\mathcal{N}=6$ Chern-Simons matter theory with gauge group $U(N)_{k}\times U(N)_{-k}$, also referred as ABJM, Wilson loops are among the most interesting observables. In this theory circular $\frac{1}{6}$-BPS bosonic \cite{Drukker:2008zx,Chen:2008bp,Rey:2008bh} and $\frac{1}{2}$-BPS fermionic \cite{Drukker:2009hy} Wilson loops were constructed and their expectation values calculated in the planar limit at strong coupling using localization \cite{Marino:2009jd}. Planar corrections to these Wilson loops have been studied in detail \cite{Drukker:2010nc}, leading to exact results at all orders in $1/N$ for any winding through the Fermi gas approach \cite{Klemm:2012ii,Okuyama:2016deu}. Latitude Wilson loops in ABJM were constructed in \cite{Cardinali:2012ru} and depend effectively on a parameter $\nu\in[0,1]$ which for the fermionic case interpolates between the $\frac{1}{2}$-BPS circular Wilson loop ($\nu=1$) and a $\frac{1}{6}$-BPS latitude Wilson loop with trivial expectation value ($\nu=0$). These latitude Wilson loops have been extensively studied in perturbation theory in \cite{Bianchi:2014laa} and recently a result for their expectation value at all orders in the coupling was proposed in \cite{Bianchi:2018bke}.

From the perspective of string theory in $AdS_{4}\times CP^{3}$, much less is known regarding the Wilson loop sector. The classical string configurations dual to bosonic Wilson loops are not known, while for fermionic Wilson loops the dual string solutions have been found for the circle \cite{Drukker:2008zx} and latitude \cite{Correa:2014aga}. At strong coupling in the string theory side, the leading contribution comes from the regularized minimal area and matches the field theory predictions. However, string theory computations in $AdS_{4}\times CP^{3}$ have been so far unsuccessful beyond leading order. In \cite{Kim:2012nd} the semiclassical string partition function dual to the $\frac{1}{2}$-BPS fermionic circular Wilson loop was computed using the Gel'fand-Yaglom method, leading to a mismatch with the localization result found on \cite{Marino:2009jd}. Recently, the computation of the ratio of fermionic $\frac{1}{2}$-BPS circular and $\frac{1}{6}$-BPS latitude semiclassical string partition functions using zeta function techniques lead to discrepancies with the prediction from \cite{Bianchi:2018bke}, although a perturbative heat kernel approach agrees at first order for small latitude angle \cite{Aguilera-Damia:2018bam}. 

In this paper we revisit 1-loop corrections to holographic Wilson loops in $AdS_{4}\times CP^{3}$ using the techniques developed in \cite{Cagnazzo:2017sny,Medina-Rincon:2018wjs}. First, we consider the ratio of semiclassical string partitions functions corresponding to the $\frac{1}{2}$-BPS circular and $\frac{1}{6}$-BPS latitude fermionic Wilson loops. For this we explicitly evaluate the functional determinants involved using phaseshifts and diffeomorphism invariant regulators in a calculation analogous to \cite{Cagnazzo:2017sny}. As we will see, due to the presence of fermionic massless operators in the $AdS_{4}\times CP^{3}$ case, the choice of fermionic boundary conditions has to be revised. Then, following the logic of \cite{Medina-Rincon:2018wjs}, we consider the 1-loop string theory computation of the $\frac{1}{2}$-BPS fermionic circular Wilson loop by considering its ratio with a configuration dual to the $\frac{1}{6}$-BPS latitude with trivial expectation value. The later result follows directly from the first computation and is considerably simpler than in the $AdS_{5}\times S^{5}$ case \cite{Medina-Rincon:2018wjs} where the relative number of zero modes in the ratio was non-zero.\\

This paper is organized as follows. In section \ref{sec2} we review the main field theory predictions for latitude and circular Wilson loops in ABJM, as well as the corresponding classical string solutions. Section \ref{sec3} describes the setup for the calculation of the semiclassical string partition functions. Then, in section \ref{sec4}, the functional determinants are evaluated explicitly and the corresponding boundary conditions are discussed. Section \ref{sec5}  collects all the pieces entering the ratio of $\frac{1}{6}$-BPS latitude and $\frac{1}{2}$-BPS circular Wilson loops, while section \ref{sec6} discusses the 1-loop string theory result for the fermionic $\frac{1}{2}$-BPS circle. Conclusions and open problems are presented in section \ref{sec7}.


\section{The circular and latitude Wilson loops}\label{sec2}
\subsection{Latitude Wilson loops in ABJM}

The family of fermionic latitude Wilson loops in ABJM constructed in \cite{Cardinali:2012ru} are parametrized by an effective parameter $\nu=\sin2\alpha_{\text{int}}\cos\theta_0\in[0,1]$, where $\alpha_{\text{int}}\in[0,\frac{\pi}{2}]$ can be arbitrarily chosen and  denotes an angle that governs the coupling of matter to the internal R-symmetry space, while $\theta_{0}$ is a geometric angle parameterizing the latitude contour. The expectation value for this family of Wilson loops at non-integer framing $f=\nu$ was obtained in \cite{Bianchi:2018bke} resulting in\footnote{In this expression we have omitted the overall normalization factor $R = {\left( {{e^{ - i\frac{{\pi \nu }}{2}}} - {e^{i\frac{{\pi \nu }}{2}}}} \right)^{ - 1}}$ used by the authors of \cite{Bianchi:2018bke} such that at weak coupling the expectation value would go as $\left\langle W \right\rangle  \sim 1 + \mathcal{O}\left(\lambda\right)$. The normalization used here, where the Wilson loop expectation value does not start with one at weak coupling, is common in Chern-Simons theory \cite{Drukker:2010nc}. }
\begin{align}\label{italianresult}
{\left\langle {{W_F^{\frac{1}{6}}}\left( \nu  \right)} \right\rangle _\nu } = \frac{{i\nu \Gamma \left( { - \frac{\nu }{2}} \right)\csc \left( {\frac{{2\pi \nu }}{k}} \right)\text{Ai}\left( {{{\left( {\frac{2}{{{\pi ^2}k}}} \right)}^{ - 1/3}}\left( {N - \frac{k}{{24}} - \frac{{6\nu  + 1}}{{3k}}} \right)} \right)}}{{{2^{\nu  + 1}}\sqrt \pi  \Gamma \left( {\frac{{3 - \nu }}{2}} \right)\csc \left( {\frac{{\pi \nu }}{2}} \right)\text{Ai}\left( {{{\left( {\frac{2}{{{\pi ^2}k}}} \right)}^{ - 1/3}}\left( {N - \frac{k}{{24}} - \frac{1}{{3k}}} \right)} \right)}}.
\end{align}
By making use of the relations
\begin{align}
\lambda=\frac{N}{k} = \frac{{{{\log }^2}\kappa }}{{2{\pi ^2}}} + \frac{1}{{24}}+\mathcal{O}\left( {{\kappa ^{ - 2}}} \right), &&
{g_s} = \frac{{2\pi i}}{k},
\end{align}
equation \eqref{italianresult} can be expanded at strong coupling in the genus series
\begin{align}
{\left\langle {W_F^{\frac{1}{6}}\left( \nu  \right)} \right\rangle _\nu } = \sum\limits_{g = 0}^\infty  {g_s^{2g - 1}{{\left. {{{\left\langle {W_F^{\frac{1}{6}}\left( \nu  \right)} \right\rangle }_\nu }} \right|}_g}} ,
\end{align}
where $g_{s}>0$ denotes the string coupling.
At genus 0, the Wilson loop expectation value reduces to
\begin{align}\label{wlatitude}
{\left. {{{\left\langle {{W_F^{\frac{1}{6}}}\left( \nu  \right)} \right\rangle }_\nu }} \right|_{g = 0}} =  - \frac{{{2^{ - \nu  - 1}}{\kappa ^\nu }\Gamma \left( { - \frac{\nu }{2}} \right)\sin \left( {\frac{{\pi \nu }}{2}} \right)}}{{\sqrt \pi  \Gamma \left( {\frac{3}{2} - \frac{\nu }{2}} \right)}}.
\end{align}
Of special interest is the case when $\nu=1$, where the latitude reduces to the $\frac{1}{2}$-BPS circular configuration. In this case the Wilson loop expectation value at strong coupling is given by
\begin{align}\label{wcircle}
\left\langle {W_F^{\frac{1}{2}}} \right\rangle=\mathop {\lim }\limits_{\nu  \to 1} {\left. {{{\left\langle {{W_F^{\frac{1}{6}}}\left( \nu  \right)} \right\rangle }_\nu }} \right|_{g = 0}} =  \frac{{{1}}}{2}\, e^{\pi \sqrt {2\lambda } }.
\end{align}
Another interesting case is when $\nu=0$, where the Wilson loop has trivial expectation value
\begin{align}\label{wtrivial}
 \mathop {\lim }\limits_{\nu  \to 0} {\left. {{{\left\langle {{W_F^{\frac{1}{6}}}\left( \nu  \right)} \right\rangle }_\nu }} \right|_{g = 0}} = 1
\end{align}
and it reduces to a configuration similar to the one obtained in \cite{Zarembo:2002an} for $AdS_{5}\times S^5$.

In this paper we set $\nu=\cos\theta_{0}$ for simplicity and consider the ratio of $\frac{1}{2}$-BPS circular and $\frac{1}{6}$-BPS latitude Wilson loops, for which the field theory prediction at strong coupling is
\begin{align}\label{Theprediction}
\frac{{{{\left. {{{\left\langle {{W_F^{\frac{1}{2}}}\left( 1 \right)} \right\rangle }_1}} \right|}_{g = 0}}}}{{{{\left. {{{\left\langle {{W_F^{\frac{1}{6}}}\left( \nu  \right)} \right\rangle }_\nu }} \right|}_{g = 0}}}} = {e^{\pi \sqrt {2\lambda } \left( {1 - \cos {\theta _0}} \right)}} \frac{{\Gamma \left( {2 + \cos {\theta _0}} \right)\Gamma \left( {\frac{{3 - \cos {\theta _0}}}{2}} \right)}}{{2\Gamma \left( {\frac{{3 + \cos {\theta _0}}}{2}} \right)}}.
\end{align}
Later in section \ref{sec6}, we study the case in which the latitude has trivial expectation value as in \eqref{wtrivial}, recovering the result for the fermionic circular Wilson loop \eqref{wcircle}.


\subsection{Classical strings in $AdS_{4}\times CP^{3}$}
We briefly present the field content of the $AdS_{4}\times CP^{3}$  background before reviewing the string configuration dual to the femionic latitude Wilson loop.
The metric for $CP^{3}$ is given by
\begin{align}
ds_{C{P^3}}^2 = \frac{1}{4}\left[ d{\alpha ^2} + {{\cos }^2}\frac{\alpha }{2}\left( {d\theta _1^2 + {{\sin }^2}{\theta _1}d\varphi _1^2} \right) + {{\sin }^2}\frac{\alpha }{2}\left( {d\theta _2^2 + {{\sin }^2}{\theta _2}d\varphi _2^2} \right)\right.\nonumber\\
\left. + {{\cos }^2}\frac{\alpha }{2}{{\sin }^2}\frac{\alpha }{2}{{\left( {d\chi  + \cos {\theta _1}d{\varphi _1} - \cos {\theta _2}d{\varphi _2}} \right)}^2} \right],
\end{align}
where the angles have the range $0 \le \alpha ,{\theta _1},{\theta _2} \le \pi $, while $0 \le {\varphi _1},{\varphi _2} \le 2\pi$ and $0 \le \chi  \le 4\pi $.
The metric for $AdS_{4}$ is
\begin{align}
ds_{Ad{S_4}}^2 = \frac{{dx_\mu ^2 + d{z^2}}}{{{z^2}}},
\end{align}
while the full metric is given by
\begin{align}
d{s^2} = ds_{Ad{S_4}}^2 + 4ds_{C{P^3}}^2 .
\end{align}
This type IIA background is additionally supported by the fields
\begin{align}
\Phi  = \ln \frac{{2L}}{k}, && {F_2} = \frac{k}{4}d{A},&& {F_4} = \frac{3}{2}k{L^2}\ {\rm{Vol}}\left( {Ad{S_4}} \right),
\end{align}
where ${L^2} = \pi \sqrt {2\lambda }$ and $A$ is a 1-form given by
\begin{align}
A = \cos \alpha d\chi  + 2{\cos ^2}\frac{\alpha }{2}\cos {\theta _1}d{\varphi _1} + 2{\sin ^2}\frac{\alpha }{2}\cos {\theta _2}d{\varphi _2}.
\end{align}

The classical string solution dual to the fermionic latitude Wilson loop in ABJM is given by \cite{Correa:2014aga}
\begin{align}
{x^\mu } &= \left\{ {\frac{{\cos \tau }}{{\cosh \sigma }},\frac{{\sin \tau }}{{\cosh \sigma }},0} \right\},   & z &= \tanh \sigma,\nonumber\\
\alpha &= 0,\qquad\qquad\qquad\qquad\qquad  \varphi_{1}=\tau,\qquad\qquad  &\sin {\theta _1} &= \frac{1}{{\cosh \left( {\sigma  + {\sigma _0}} \right)}},
\end{align}
where $\sigma_{0}$ is related to the latitude angle $\theta_{0}$ by
\begin{equation}
\cos {\theta _0} = \tanh {\sigma _0}.
\end{equation}
It is easy to see that the string extends in the bulk of $AdS$ and ends in a circle at the boundary $\sigma=0$, while in $CP^{3}$ the angle $\theta_{1}\to\theta_{0}$ when $\sigma\to 0$.


\section{Setup}\label{sec3}

In the gauge-string correspondence the string theory dual to the Wilson loop expectation value corresponds to the string partition function. The later can be evaluated semiclassically by gauge fixing both $\kappa$-symmetry and the worldsheet metric, and considering small fluctuations around the classical string solution \cite{Drukker:2000ep}. Expansion of the Green-Schwarz action to second order in fluctuations reduces the path integral to a Gaussian integral, which after integration amounts to the evaluation of functional determinants. Schematically this reduces to
\begin{align}\label{WZ}
\left\langle W \right\rangle  ={Z_{{\rm{string}}}}\mathop  = \limits^{\lambda  \to  \infty} {e^{ - S\left[ {{X_{{\rm{cl}}}}} \right]}}\ \frac{{{{\det }^{1/2}}\mathcal{D}}}{{{{\det }^{1/2}}\mathcal{K}}},
\end{align}
where the exponential contains the classical piece, while the semiclassical piece is given in terms of determinants of transversal bosonic and fermionic operators $\mathcal{K}$  and $\mathcal{D}$, respectively. 

In expression \eqref{WZ} we have omitted the contribution from the ratio of ghost Faddev-Popov determinants and bosonic directions longitudinal to the worldsheet. Despite ghost and longitudinal modes having the same differential operator, the determinants are not necessarily the same due to different boundary conditions \cite{Drukker:2000ep}. The omission of these contributions is justified when considering ratios of Wilson loops with the same topology as possible contributions are expected to cancel in the ratio. Note that this was the case for ratios of circular Wilson loops in $AdS_{5}\times S^{5}$ \cite{Cagnazzo:2017sny,Medina-Rincon:2018wjs}. 

Following the formalism of \cite{Drukker:2000ep}, we fix the worldsheet metric to be the metric induced by the classical solution
\begin{align}\label{WSmetric}
d{s^2_{\text{ws}}} = {\Omega ^2}\left( {d{\tau ^2} + d{\sigma ^2}} \right),&&
{\Omega ^2} = \frac{1}{{{{\sinh }^2}\sigma }} + \frac{1}{{{{\cosh }^2}\left( {\sigma  + {\sigma _0}} \right)}}.
\end{align}

The classical contribution to the partition function is given in terms of the regularized area
\begin{align}\label{classypiece}
S\left[ {{X_{\text{cl}}}} \right] = {\left. {\sqrt {\frac{\lambda }{2}} \int\limits_0^{2\pi } {\int\limits_\varepsilon ^\infty  {\left( {\frac{1}{{{{\sinh }^2}\sigma }} + \frac{1}{{{{\cosh }^2}\left( {\sigma  + {\sigma _0}} \right)}}} \right)} d\sigma d\tau } } \right|_{{\rm{reg}}}} =  - \pi \sqrt {2\lambda } \tanh {\sigma _0}
\end{align}
and is in perfect agreement with the exponential behaviour in \eqref{Theprediction}. We will now focus on the semiclassical piece which accounts for the factor in front of the exponential in equation \eqref{Theprediction}.

\subsection{The semiclassical partition function}
Expansion of the Green-Schwarz action to second order in fluctuations around the latitude solution results in the following ratio of determinants \cite{Aguilera-Damia:2018bam}
\begin{align}\label{TheZ}
Z = \frac{{\det {\mathcal{D}_{1 + }}\, \det {\mathcal{D}_{1 - }}\, {{\det }^{1/2}}{\mathcal{D}_{2 + }}\, {{\det }^{1/2}}{\mathcal{D}_{2 - }}\, {{\det }^{1/2}}{\mathcal{D}_{3 + }}{{\det }^{1/2}}\, {\mathcal{D}_{3 - }}}}{{\det {\mathcal{K}_1}\, \det {\mathcal{K}_{2 + }}\, \det {\mathcal{K}_{2 - }}\, {{\det }^{1/2}}\, {\mathcal{K}_{3 + }}\, {{\det }^{1/2}}{\mathcal{K}_{3 - }}}},
\end{align}
where the untilded operators above are given in terms of tilded operators through
\begin{align}
\mathcal{K} = \frac{1}{{{\Omega ^2}}}\widetilde{\mathcal{K}}, && \mathcal{D} = \frac{1}{{{\Omega ^{3/2}}}}\widetilde{\mathcal{D}}{\Omega ^{1/2}},
\end{align}
and the tilded operators are
\begin{align}
{\widetilde{\mathcal{K}}_1} &=  - \partial _\tau ^2 - \partial _\sigma ^2 + \frac{2}{{{{\sinh }^2}\sigma }},\\
{\widetilde{\mathcal{K}}_{2\alpha}} &=  - \partial _\tau ^2 - \partial _\sigma ^2 + \alpha i\left( {\tanh \left( {\sigma  + {\sigma _0}} \right) - 1} \right){\partial _\tau }\nonumber\\
 &\qquad\qquad\qquad\qquad+ \frac{1}{4}\left( {\tanh \left( {\sigma  + {\sigma _0}} \right) - 1} \right)\left( {1 + 3\tanh \left( {\sigma  + {\sigma _0}} \right)} \right),\\
{\widetilde{\mathcal{K}}_{3\alpha }} &=  - \partial _\tau ^2 - \partial _\sigma ^2 + 2\alpha i\left( {\tanh \left( {2\sigma  + {\sigma _0}} \right) - 1} \right){\partial _\tau }\nonumber \\
&\qquad\qquad\qquad\qquad+ \left( {\tanh \left( {2\sigma  + {\sigma _0}} \right) - 1} \right)\left( {1 + 3\tanh \left( {2\sigma  + {\sigma _0}} \right)} \right),
\end{align}
\begin{align}
{\widetilde{\mathcal{D}}_{1\alpha }}& = i{\tau _1}{\partial _\sigma } - {\tau _2}\left[ {i{\partial _\tau } + \frac{\alpha }{2}\left( {1 - \tanh \left( {2\sigma  + {\sigma _0}} \right)} \right)} \right] + \frac{1}{{\Omega {{\sinh }^2}\sigma }}{\tau _3} + \frac{\alpha }{{\Omega {{\cosh }^2}\left( {\sigma  + {\sigma _0}} \right)}},\\
\widetilde{\mathcal{D}}_{2\alpha } &= i{\tau _1}{\partial _\sigma } - {\tau _2}\left[ {i{\partial _\tau } + \frac{\alpha }{2}\left( {\tanh \left( {\sigma  + {\sigma _0}} \right) - \tanh \left( {2\sigma  + {\sigma _0}} \right)} \right)} \right]\nonumber\\
&\qquad\qquad\qquad\qquad\qquad - \frac{1}{2}\left( {\frac{1}{{\Omega {{\sinh }^2}\sigma }} + \Omega } \right){\tau _3} - \frac{\alpha }{{2\Omega {{\cosh }^2}\left( {\sigma  + {\sigma _0}} \right)}},\\
{\widetilde{\mathcal{D}}_{3\alpha }} &= i{\tau _1}{\partial _\sigma } - {\tau _2}\left[ {i{\partial _\tau } + \frac{\alpha }{2}\left( {2 - \tanh \left( {\sigma  + {\sigma _0}} \right) - \tanh \left( {2\sigma  + {\sigma _0}} \right)} \right)} \right]\nonumber\\
&\qquad\qquad\qquad\qquad\qquad - \frac{{\alpha\, \text{sech}^2\left( {\sigma  + {\sigma _0}} \right)}}{{2\Omega }}\left({\mathbbm{1} - \alpha {\tau _3}} \right), \end{align}
with the variable $\alpha$ being either a $+$ or $-$ sign and $\tau_{i}$ denoting the Pauli matrices. The operator $\mathcal{K}_{1}$ corresponds to two fluctuations directions along $AdS_{4}$, $\mathcal{K}_{2\alpha}$ comes from four fluctuation modes along $CP^{3}$ while $\mathcal{K}_{3\alpha}$ results from a mixing of the remaining transversal fluctuations along $AdS_{4}$ and $CP^{3}$. Meanwhile, the fermionic operators $\mathcal{D}_{i}$ result from $\kappa$-symmetry gauge-fixing and choosing an appropriate basis for the gamma matrices entering the fermionic kinetic term in the Green-Schwarz action.

The asymptotic behaviour of the untilded operators far from the boundary is such that
\begin{align}
\mathop {\lim }\limits_{\sigma  \to \infty } {{\widetilde {\mathcal K}}_i} &= \widetilde{\mathcal{K}}_{\infty}=  - \partial _\tau ^2 - \partial _\sigma ^2 &\forall i\in\{1,2,3\},\\
\mathop {\lim }\limits_{\sigma  \to \infty } {\widetilde{\mathcal{D}}_{i \alpha}} &=\widetilde{\mathcal{D}}_{\infty}= i{\tau _1}{\partial _\sigma } - i{\tau _2}{\partial _\tau } &\forall i\in\{1,2,3\},
\end{align}
where the bosonic and fermionic asymptotic operators are related through
\begin{align}
{\left( {{{\widetilde{\mathcal D}}_\infty }} \right)^2} = \mathbbm{1}\, {\widetilde{\mathcal{K}}_\infty }.
\end{align}

\subsection{Conformal frame and invariant regulators}\label{cutoff}

To evaluate the semiclassical contribution \eqref{TheZ} we need to evaluate the corresponding functional determinants coming from string fluctuations. The later are naturally normalized with respect to the induced worldsheet metric \eqref{WSmetric} \cite{Drukker:2000ep,Cagnazzo:2017sny}
\begin{align}
\langle
 \mathrel{ \phi _1 | \phi _2 }
\rangle  = \int {d{\sigma ^2}\sqrt h }\, \phi _1^\dag {\phi _2} = \int {d\tau d\sigma\, {\Omega ^2}} \phi _1^\dag {\phi _2}
\end{align}
and the untilded operators are Hermitian with respect to this inner product.

Just as done in \cite{Cagnazzo:2017sny,Medina-Rincon:2018wjs}, for simplicity we will work with the tilded operators instead of the untilded ones appearing originally in the partition function. The tilded operators are Hermitian with respect to the inner product
\begin{align}
\widetilde{\langle{\phi _1}\, |\, {\phi _2}\rangle}= \int {d\tau d\sigma } \phi _1^\dag {\phi _2}.
\end{align}
Tilded and untilded operators are related by a conformal transformation of the form $h_{\mu\nu}=\Omega^{2}\delta_{\mu\nu}\to\delta_{\mu\nu}$. As discussed in \cite{Cagnazzo:2017sny}, this map is singular at $\sigma=\infty$ and changes the worldsheet topology from that of a disc to a semi-infinite cylinder. Evaluation of functional determinants on the cylinder requires an IR regularization implicit in our calculations of section \ref{sec4}, where artificial boundary conditions are imposed at a large but finite value $\sigma=R$. Imposing this cut-off $R$ effectively affects the determinants as in the original frame it amounts to introducing a small hole in the disk. As done in \cite{Cagnazzo:2017sny}, we will introduce a diffeomorphism-invariant regularization by choosing the area removed $s$ to be the same for all values of $\sigma_{0}$
\begin{align}
s = \int\limits_R^\infty  {\int\limits_0^{2\pi } {{\Omega ^2}d\tau d\sigma } }  \simeq 4\pi \left( {1 + {e^{ - 2{\sigma _0}}}} \right){e^{ - 2R}}.
\end{align}
Solving for $R$ one obtains the $\sigma_{0}$ dependent cut-off
\begin{align}\label{IRcutoff}
R\left( {{\sigma _0}} \right) = \frac{1}{2}\ln \frac{{8\pi }}{{s\left( {1 + \tanh {\sigma _0}} \right)}}.
\end{align}

We will now proceed to evaluate the determinants on the cylinder, or equivalently, the determinants for the tilded operators.

\section{Evaluation of determinants}\label{sec4}

Before evaluating individual determinants, we briefly present the general prescription used. For more details we refer the reader to \cite{Cagnazzo:2017sny}.

Let $\widetilde{\mathcal{K}}$ a differential operator of the form 
\begin{align}
\widetilde{\mathcal{K}}=-\partial_{\tau}^{2}-\partial_{\sigma}^{2}+V(\partial_{\tau},\sigma),
\end{align}
with $V(\partial_{\tau},\sigma)\to0$ as $\sigma\to\infty$. Fourier expansion of the eigenfunction $\phi\left(\tau,\sigma\right)=e^{-i\omega\tau}\chi\left(\sigma\right)$ along the $\tau$ direction effectively reduces the spectral problem to a 1-dimensional problem
\begin{align}\label{SpectralProblem}
\widetilde{\mathcal{K}}\, \chi  = \left[ {{\omega ^2} - \partial _\sigma ^2 + V\left( { - i\omega ,\sigma } \right)} \right]\chi  = \left( {{\omega ^2} + {p^2}} \right)\chi ,
\end{align}
with eigenvalue $\omega^{2}+p^2$ and eigenfunction $\chi$.

For bosonic operators we impose Dirichlet  boundary conditions along the $\sigma$ direction
\begin{align}\label{bBC}
\chi\left(\sigma=0\right)=0.
\end{align} 
The eigenfunctions of the asymptotic operator $\widetilde{\mathcal{K}}_{\infty}$ with vanishing potential correspond to plane waves of the form $\chi_{\infty}\propto \sin(p\sigma)$. Meanwhile, for the original operator $\widetilde{\mathcal{K}}$ the eigenfunction far from the origin approaches a plane wave solution with a scattering phaseshift due to the potential $V$
\begin{align}
\mathop {\lim }\limits_{\sigma  \to \infty } \chi  \propto  \sin \left( {p\sigma  + \delta\left(\omega,p\right) } \right).
\end{align}
To evaluate the determinant we impose an ``artificial" boundary condition $\chi(R)=0$ at the large but finite IR cut-off $R$ discussed in section \ref{cutoff}. The later results in 
\begin{align}
pR+\delta\left(\omega,p\right)=\pi n
\end{align}
for $n\in\mathbb{Z}$. The above implies a density of states for the continuum spectrum given by $\frac{dn}{dp}=\frac{R}{\pi}+\frac{\partial_{p}\delta\left(\omega,p\right)}{\pi}$. In terms of the phaseshift and the IR cut-off, the functional determinant is given by 
\begin{align}
\ln \det \widetilde{\mathcal{K}} = \sum\limits_\omega  {\int\limits_0^\infty  {\frac{{dp}}{\pi }\left( {R + {\partial _p}\delta \left( {\omega ,p} \right)} \right)\ln \left( {{\omega ^2} + {p^2}} \right)} } .
\end{align}
Notice that in the equation above only the derivative of the phaseshift plays a role, thus, phaseshifts are fixed up to a constant, which for convenience we set such that $\delta \mathop  \to \limits^{p \to \infty } 0$.

Summation over periodic Matsubara frequencies $\omega\in\mathbb{Z}$ and contour integration as shown in \cite{Cagnazzo:2017sny} result in
\begin{align}\label{lndetKB}
\ln \det \widetilde{\mathcal{K}} =  - \int\limits_0^\infty  {dp\coth \pi p\, \left[ {{\delta _ + }\left( p \right) + {\delta _ - }\left( p \right) + 2Rp} \right]} 
\end{align}
for bosons, where $\delta_{\pm}\left( p\right)=\delta\left(\pm i p,p\right)$ as integration over $\omega$ in intermediary steps\footnote{See \cite{Cagnazzo:2017sny} for more details.}  picks up poles at $\omega=\pm i p$. Thus, to compute a bosonic determinant one uses equation \eqref{lndetKB} where the phaseshift is read at large $\sigma$ after explicitly solving \eqref{SpectralProblem} with $\omega=\pm i p$
\begin{align}
{\left. {\widetilde{\mathcal{K}}\chi } \right|_{\omega  =  \pm ip}} = \left[ { - {p^2} - \partial _\sigma ^2 + V\left( { \pm p,\sigma } \right)} \right]\chi  = 0
\end{align}
subject to the boundary condition \eqref{bBC}.\\

To evaluate fermionic functional determinants, the following equation is used
\begin{align}\label{lndetKF}
\ln \det \widetilde{\mathcal{D}} =  - \int\limits_0^\infty  {dp\tanh \pi p\, \left[ {{\delta _ + }\left( p \right) + {\delta _ - }\left( p \right) + 2Rp} \right]},
\end{align}
which results from summation over anti-periodic Matsubara frequencies $\omega\in\mathbb{Z}+\frac{1}{2}$ \cite{Cagnazzo:2017sny}. After Fourier expansion of the eigenfunctions, phaseshifts are read from the oscillatory behaviour at large $\sigma$ of the 2-component spinor eigenfunctions $\chi\left(\sigma\right)$ satisfying
\begin{align}\label{fspp}
\widetilde{\mathcal{D}} \chi |_{\omega=\pm i p}=0,
\end{align}
subjected to the required boundary condition along the $\sigma$ direction.

Boundary conditions for fermions will play a key role in the present calculation. In previous works \cite{Cagnazzo:2017sny,Medina-Rincon:2018wjs,Chen-Lin:2017pay} a generic solution to the fermionic spectral problem \eqref{fspp} corresponded to a superposition of two eigenfunctions with divergent behaviour at $\sigma\to0$, namely of the form
\begin{align}\label{bcoriginal}
\mathop {\lim }\limits_{\sigma  \to 0} \chi  \simeq {A_1}{\sigma ^{ - l}}{\chi _{\left( 1 \right)}} + {A_2}{\sigma ^{ - l}}{\chi _{\left( 2 \right)}} + \mathcal{O}\left( {{\sigma ^0}} \right),
\end{align}
where $l>0$ while $A_1$ and $A_2$ denote the coefficients of the superposition. A natural choice of boundary conditions corresponded to choosing the superposition which is well behaved for $\sigma\to0$ since the resulting eigenfunction would be normalizable. Such choice of fermionic boundary conditions lead to perfect agreement with field theory predictions in \cite{Cagnazzo:2017sny,Medina-Rincon:2018wjs,Chen-Lin:2017pay}.

As we will see later, the above prescription for fermionic boundary conditions can be easily applied for the massive operators $\widetilde{\mathcal{D}}_{1\alpha}$ and $\widetilde{\mathcal{D}}_{2\alpha}$ as in these cases $l=1$. However, for the massless fermionic operator $\widetilde{\mathcal{D}}_{3\alpha}$ this prescription can not be used since the behaviour at the origin is such that $l=0$ for both eigenfunctions in the superposition. Consequently, boundary conditions for fermions have to be revised such that they are compatible with previous results, while also sorting out the ambiguity for the massless fermionic operator $\widetilde{\mathcal{D}}_{3\alpha}$.

In the cylinder, fermionic boundary conditions at $\sigma=0$ will be introduced using the projector\footnote{Note that the $\pm$ in the definition of $\Pi_{\pm}$ is not related to the $\pm$ sign resulting from the $\omega=\pm i p$ leading to \eqref{lndetKF} and \eqref{fspp}. The two are independent.}
\begin{align}
{\Pi _ \pm } = \frac{1}{2}\left( {\mathbbm{1} \pm i {\gamma _ * }{\gamma ^\mu }{n_\mu }} \right),
\end{align}
where $n_{\mu}=\{n_\tau,n_\sigma\}=\{0,1\}$ is the inward pointing unit vector orthogonal to the boundary, $\gamma^{\mu}$ denotes the gamma matrices $\gamma^{\tau}=-\tau_{2}$ and $\gamma^{\sigma}=\tau_{1}$, while $\gamma_{*}=i\gamma^{\tau}\gamma^{\sigma}=-\tau_{3}$ is the chiral matrix. It is easy to see that this projector satisfies
\begin{align}
{\Pi _ + } + {\Pi _ - } = \mathbbm{1},&& {\Pi _ \pm }{\Pi _ \pm } = {\Pi _ \pm }, && {\Pi _ \pm }{\Pi _ \mp } = 0.
\end{align}
To evaluate the functional determinants, we have performed a conformal transformation $h_{\mu\nu}=\Omega^{2}\delta_{\mu\nu}\to\delta_{\mu\nu}$ and considered a finite spatial interval $0 \le \sigma  \le R$. Appropriate boundary conditions for spinors in the finite spatial interval in flat space can be found by supersymmetry considerations and amount to two possible sets in which Dirichlet and Neumann conditions are satisfied along the different spinor components \cite{Sakai:1984vm}. By explicit calculation it can be checked that for the massive operators $\widetilde{\mathcal{D}}_{1\alpha}$ and $\widetilde{\mathcal{D}}_{2\alpha}$, the superpositions picked by the prescription of \cite{Cagnazzo:2017sny,Medina-Rincon:2018wjs,Chen-Lin:2017pay} correspond to two sets of boundary conditions 
\begin{align}\label{nuevabc}
{\left. {{\chi _ \pm }} \right|_{\sigma  = 0}} = 0\qquad \wedge\qquad  {\left. {{\partial _\sigma }{\chi _ \mp }} \right|_{\sigma  = 0}} = 0,
\end{align}
where $\chi_{\pm}=\Pi_{\pm}\chi$ denote the spinor projections. Boundary conditions of this type, where one projection satisfies Dirichlet while the complementary projection satisfies Neumann, are sometimes referred to as ``mixed" boundary conditions and are natural boundary conditions for spin-$\frac{1}{2}$ fields \cite{Fursaev:2011zz}. 

We will see that for the massless fermionic operator $\widetilde{\mathcal{D}}_{3\alpha}$ imposing a set of boundary conditions \eqref{nuevabc} fixes the superposition and its corresponding phaseshift, leading to agreement with the field theory prediction of \cite{Bianchi:2018bke}. Note that similar mixed boundary conditions for massless fermions were also necessary to match field theory predictions in the calculation of the 1-loop partition function for strings in $AdS_{4}\times CP^{3}$ ending in cusped lines at the $AdS$ boundary \cite{Aguilera-Damia:2014bqa}, solving previous discrepancies found in \cite{Forini:2012bb}.

\subsection*{Operator $\widetilde{\mathcal{K}}_{1}$}
The spectral problem for this operator is
\begin{align}
\left( { - \partial _\sigma ^2 + \frac{2}{{{{\sinh }^2}\sigma }}} \right){\chi _1} = {p^2}{\chi _1}.
\end{align}
The eigenfunctions satisfying this equation are a superposition of the Jost functions
\begin{align}
{Y_1} = {e^{ip\sigma }}\frac{{ip - \coth \sigma }}{{ip - 1}}, && {{\bar Y}_1} = {e^{ - ip\sigma }}\frac{{ip + \coth \sigma }}{{ip + 1}}.
\end{align}
The superposition satisfying the boundary condition $\chi_{1}\left(\sigma=0\right)=0$ corresponds to
\begin{align}
{\chi _1} \propto {e^{ip\sigma }}\left( {ip - \coth \sigma } \right) - {e^{ - ip\sigma }}\left( { - ip - \coth \sigma } \right).
\end{align}
Consequently, the phaseshift is given by
\begin{align}
\delta_{1}=\delta_{1}^{\pm} =-\arctan p + \frac{\pi}{2}
\end{align}
and the determinant is
\begin{align}
\ln {\det {{\widetilde{ \mathcal{K}}}_1}} &= \int\limits_0^\infty  {\coth \pi p\left( {2\arctan p}-\pi -2Rp\right)dp}.
\end{align}


\subsection*{Operator $\widetilde{\mathcal{K}}_{2\alpha}$}
The spectral problem for this operator is
\begin{align}
\left[ { - \partial _\sigma ^2 + \frac{1}{4}\left( {\tanh \left( {\sigma  + {\sigma _0}} \right) - 1} \right)\left( { \pm 4\alpha ip + 1 + 3\tanh \left( {\sigma  + {\sigma _0}} \right)} \right)} \right]{\chi _{2\alpha }} = {p^2}{\chi _{2\alpha }}.
\end{align}
The eigenfunctions satisfying this equation are a superposition of the Jost functions
\begin{align}
{Y_{2\alpha }} &= {e^{ \pm \alpha ip\sigma }}{\left( {\frac{{1 + \tanh \left( {\sigma  + {\sigma _0}} \right)}}{2}} \right)^{ - 1/2}}\left( {\frac{{ \pm \alpha ip - \frac{{1 + \tanh \left( {\sigma  + {\sigma _0}} \right)}}{2}}}{{ \pm \alpha ip - 1}}} \right),\\  
{\overline Y _{2\alpha }} &= {e^{ \mp \alpha ip\sigma }}{\left( {\frac{{1 + \tanh \left( {\sigma  + {\sigma _0}} \right)}}{2}} \right)^{1/2}}.
\end{align}
The superposition satisfying $\chi_{2\alpha}\left(\sigma=0\right)=0$ is given by
\begin{align}
{\chi _{2\alpha}} &\propto {e^{ \pm \alpha ip\sigma }}\sqrt {\frac{{1 + \tanh {\sigma _0}}}{{1 + \tanh \left( {\sigma  + {\sigma _0}} \right)}}} \left( { \pm \alpha ip - \frac{{1 + \tanh \left( {\sigma  + {\sigma _0}} \right)}}{2}} \right)\nonumber\\
&\qquad\qquad + {e^{ \mp \alpha ip\sigma }}\sqrt {\frac{{1 + \tanh \left( {\sigma  + {\sigma _0}} \right)}}{{1 + \tanh {\sigma _0}}}} \left( { \mp \alpha ip + \frac{{1 + \tanh {\sigma _0}}}{2}} \right).
\end{align}
From the above, the phaseshift is given by
\begin{align}
\delta_{2\alpha}=\delta _{2\alpha }^ \pm  = - \frac{1}{2}\arctan p + \frac{1}{2}\arctan \frac{{2p}}{{1 + \tanh {\sigma _0}}}
\end{align}
and the determinant is
\begin{align}
\ln {\det {{\widetilde{ \mathcal{K}}}_{2\alpha }}}&= \int\limits_0^\infty  {\coth \pi p\left( {\arctan p - \arctan \frac{{2p}}{{1 + \tanh {\sigma _0}}}}-2Rp \right)dp}. 
\end{align}


\subsection*{Operator $\widetilde{\mathcal{K}}_{3\alpha}$}
The spectral problem for this operator is
\begin{align}
\left[ { - \partial _\sigma ^2 + \left( {\tanh \left( {2\sigma  + {\sigma _0}} \right) - 1} \right)\left( { \pm 2\alpha ip + 1 + 3\tanh \left( {2\sigma  + {\sigma _0}} \right)} \right)} \right]{\chi _{3\alpha }} = {p^2}{\chi _{3\alpha }}.
\end{align}
The eigenfunctions satisfying the equation above are a superposition of the Jost functions
\begin{align}
{Y_{3\alpha }} &= {e^{ \pm \alpha ip\sigma }}{\left( {\frac{{1 + \tanh \left( {2\sigma  + {\sigma _0}} \right)}}{2}} \right)^{ - 1/2}}\left( {\frac{{ \pm \alpha ip - 1 - \tanh \left( {2\sigma  + {\sigma _0}} \right)}}{{ \pm \alpha ip - 2}}} \right),\\
{\overline Y _{3\alpha }} &= {e^{ \mp \alpha ip\sigma }}{\left( {\frac{{1 + \tanh \left( {2\sigma  + {\sigma _0}} \right)}}{2}} \right)^{1/2}}.
\end{align}
The superposition satisfying the boundary condition $\chi_{3\alpha}\left(\sigma=0\right)=0$ is given by
\begin{align}
{\chi _{3\alpha}} &\propto {e^{ \pm \alpha ip\sigma }}\sqrt {\frac{{1 + \tanh {\sigma _0}}}{{1 + \tanh \left( {2\sigma  + {\sigma _0}} \right)}}} \left( { \pm \alpha ip - 1 - \tanh \left( {2\sigma  + {\sigma _0}} \right)} \right)\nonumber\\
&\qquad\qquad + {e^{ \mp \alpha ip\sigma }}\sqrt {\frac{{1 + \tanh \left( {2\sigma  + {\sigma _0}} \right)}}{{1 + \tanh {\sigma _0}}}} \left( { \mp \alpha ip + 1 + \tanh {\sigma _0}} \right).
\end{align}
From the above, the phaseshift is given by
\begin{align}
\delta_{3\alpha}=\delta _{3\alpha }^ \pm  = - \frac{1}{2}\arctan \frac{p}{2} + \frac{1}{2}\arctan \frac{p}{{1 + \tanh {\sigma _0}}} 
\end{align}
and the determinant is
\begin{align}
\ln {\det {{\widetilde{ \mathcal{K}}}_{3\alpha }}} &= \int\limits_0^\infty  {\coth \pi p\left( {\arctan \frac{p}{2} - \arctan \frac{p}{{1 + \tanh {\sigma _0}}}}-2Rp \right)dp} .
\end{align}
%

\subsection*{Operator $\widetilde{\mathcal{D}}_{1\alpha}$}
The spectral problem for this operator can be written as
{\small{\begin{align}
\left[ {{\tau _3}{\partial _\sigma } + \frac{i}{{\Omega\, {{\sinh }^2}\sigma }}{\tau _1} + \frac{\alpha }{{\Omega\, {{\cosh }^2}\left( {\sigma  + {\sigma _0}} \right)}}{\tau _2} - \frac{\alpha }{2}\left( {1 - \tanh \left( {2\sigma  + {\sigma _0}} \right)} \right)\mathbbm{1}} \right]{\chi _{1\alpha }} =  \pm ip\, {\chi _{1\alpha }}.
\end{align}}}
The eigenfunctions satisfying this equation are given by a superposition of
\begin{align}
{Y _{1 \alpha }} &= {e^{ \pm i\alpha p\sigma }}\left( {{\delta _{\alpha ,+}}\left[ {\begin{array}{*{20}{c}}
  {c_1^{\text I}} \\ 
  {c_1^{\text{II}}} 
\end{array}} \right] + {\delta _{\alpha , -}}\left[ {\begin{array}{*{20}{c}}
  {c_1^{\text{II}}} \\ 
  {c_1^{\text{I}}} 
\end{array}} \right]} \right),\\
{\bar Y _{1 \alpha }} &= {e^{ \mp i\alpha p\sigma }}\left( {{\delta _{\alpha , + }}\left[ {\begin{array}{*{20}{c}}
  {\bar c_1^{\text{I}}} \\ 
  {\bar c_1^{{\text{II}}}} 
\end{array}} \right] + {\delta _{\alpha , - }}\left[ {\begin{array}{*{20}{c}}
  {\bar c_1^{{\text{II}}}} \\ 
  {\bar c_1^{\text{I}}} 
\end{array}} \right]} \right),
\end{align}
where
{\small{\begin{align}
c_1^{\text I} &= \frac{{{e^{\sigma /2}}{e^{{\sigma _0}{\text{/4}}}}\left( { \pm i\alpha p - \frac{1}{2}\left( {\frac{{\cosh \left( {2\sigma  + {\sigma _0}} \right)}}{{\sinh \sigma {\kern 1pt} \,\cosh \left( {\sigma  + {\sigma _0}} \right)}} - 1} \right)} \right)}}{{\left( { \pm i\alpha p - \frac{1}{2}} \right){2^{1/4}}{{\cosh }^{1/4}}\left( {2\sigma  + {\sigma _0}} \right)}},\\
c_1^{\text{II}} &= \frac{{i\alpha {e^{\sigma /2}}{e^{{\sigma _0}{\text{/4}}}}\Omega }}{{\left( { \pm i\alpha p - \frac{1}{2}} \right){2^{5/4}}{{\cosh }^{1/4}}\left( {2\sigma  + {\sigma _0}} \right)}},\\
\bar c_1^{\text I} &= \frac{{i\alpha {2^{ - 7/4}}{e^{ - \sigma /2}}{e^{ - {\sigma _0}/4}}\Omega }}{{\left( { \pm i\alpha p + \frac{3}{2}} \right)\left( { \pm i\alpha p + \frac{1}{2}} \right)\cosh {\sigma _0}{{\cosh }^{3/4}}\left( {2\sigma  + {\sigma _0}} \right)}}\times\nonumber\\
&\left( {\left( { \pm i\alpha p + \frac{1}{2}} \right)\left( {2 + \cosh \left( {2\left( {\sigma  + {\sigma _0}} \right)} \right) - \cosh 2\sigma } \right) + \sinh \left( {2\left( {\sigma  + {\sigma _0}} \right)} \right) - \sinh {\text{2}}\sigma } \right),\\
\bar c_1^{\text{II}} &= \frac{{{2^{ - 7/4}}{e^{\sigma /2}}{e^{ - 5{\sigma _0}/4}}{\Omega ^{1/2}}}}{{\left( { \pm i\alpha p{\text{ + }}\frac{3}{2}} \right)\left( { \pm i\alpha p{\text{ + }}\frac{1}{2}} \right){{\cosh }^{1/4}}{\sigma _0} \sqrt {\sinh \sigma \cosh \left( {\sigma  + {\sigma _0}} \right)} }}\times\nonumber\\
&\left[ {{e^{ - 3\sigma }}\left( {{p^2} + \frac{1}{4}} \right) + \left( { \pm i\alpha p + \frac{3}{2}} \right)\left( {{e^{ - \sigma }}\left( { \pm i\alpha p - \frac{1}{2}} \right) + 2{e^{2{\sigma _0}}}\sinh \sigma \left( { \pm i\alpha p{\text{ + }}\frac{{\coth \sigma }}{2}} \right)} \right)} \right].
\end{align}}}
The asymptotic behaviour at large $\sigma$ of these solutions is given by
\begin{align}
\mathop {\lim }\limits_{\sigma  \to \infty } {Y _{1 \alpha }} &= {e^{ \pm i\alpha p\sigma }}\left( {{\delta _{\alpha , + }}\left[ {\begin{array}{*{20}{c}}
  1 \\ 
  0 
\end{array}} \right] + {\delta _{\alpha , - }}\left[ {\begin{array}{*{20}{c}}
  0 \\ 
  1 
\end{array}} \right]} \right),\\
\mathop {\lim }\limits_{\sigma  \to \infty } {{\bar Y }_{1 \alpha }} &= {e^{ \mp i\alpha p\sigma }}\left( {{\delta _{\alpha , + }}\left[ {\begin{array}{*{20}{c}}
  0 \\ 
  1 
\end{array}} \right] + {\delta _{\alpha , - }}\left[ {\begin{array}{*{20}{c}}
  1 \\ 
  0 
\end{array}} \right]} \right),
\end{align}
while for $\sigma\to0$ they have the following behaviour
\begin{align}
\mathop {\lim }\limits_{\sigma  \to 0} {Y _{1 \alpha }} = \frac{{{v_1}}}{\sigma }\left( {{\delta _{\alpha , + }}\left[ {\begin{array}{*{20}{c}}
  1 \\ 
  { - \alpha i} 
\end{array}} \right] + {\delta _{\alpha , - }}\left[ {\begin{array}{*{20}{c}}
  { - \alpha i} \\ 
  1 
\end{array}} \right]} \right) + \mathcal{O}\left( {{\sigma ^0}} \right),\\
\mathop {\lim }\limits_{\sigma  \to 0} {{\bar Y }_{1 \alpha }} = \frac{{{{\bar v}_1}}}{\sigma }\left( {{\delta _{\alpha , + }}\left[ {\begin{array}{*{20}{c}}
  1 \\ 
  { - \alpha i} 
\end{array}} \right] + {\delta _{\alpha , - }}\left[ {\begin{array}{*{20}{c}}
  { - \alpha i} \\ 
  1 
\end{array}} \right]} \right) + \mathcal{O}\left( {{\sigma ^0}} \right),
\end{align}
with
\begin{align}
{v_1} =  - \frac{{{{\left( {1 + \tanh {\sigma _0}} \right)}^{1/4}}}}{{{2^{5/4}}\left( { \pm i\alpha p - \frac{1}{2}} \right)}},&&
{\bar v_1} = \frac{{i\alpha {2^{ - 3/4}}\left( { \pm i\alpha p + \frac{1}{2} + \tanh {\sigma _0}} \right)}}{{{{\left( {1 + \tanh {\sigma _0}} \right)}^{1/4}}\left( { \pm i\alpha p + \frac{3}{2}} \right)\left( { \pm i\alpha p + \frac{1}{2}} \right)}}.\end{align}
A natural choice of boundary conditions is to pick the superposition which is well behaved at $\sigma\to0$. This fixes the superposition up to an overall constant
\begin{align}
{\chi _{1 \alpha }} \propto {Y_{1 \alpha }}{\bar v_1} - {\bar Y_{1 \alpha }}{v_1}.
\end{align}
Note that this superposition satisfies for both $\pm$ signs coming from $\omega=\pm i p$ and $\alpha$
\begin{align}
{\left. {{\Pi _ + }{\chi _{1\alpha }}} \right|_{\sigma  = 0}} = 0, && {\left. {{\Pi _ - }{\partial _\sigma }{\chi _{1\alpha }}} \right|_{\sigma  = 0}} = 0.
\end{align}
From the superposition, the phaseshift is given by
{\small{\begin{align}
{\delta _{1\, \alpha }} = \delta _{1\, \alpha }^ \pm  =\pm\frac{\alpha}{2}{\rm{Arg}}\left(\frac{\bar{v}_{1}}{v_{1}}\right)=- \arctan 2p - \frac{1}{2}\arctan \frac{{2p}}{3}+\frac{1}{2}\arctan \frac{p}{{\frac{1}{2} + \tanh {\sigma _0}}} + \frac{\pi}{2}
\end{align}}}
%
and the determinant is
{\small{\begin{align}
\ln {\det {{\widetilde{ \mathcal{D}}}_{1\alpha }}}&= \int\limits_0^\infty  {\tanh \pi p\left( {2\arctan 2p + \arctan \frac{{2p}}{3} - \arctan \frac{p}{{\frac{1}{2} + \tanh {\sigma _0}}}} -\pi-2Rp\right)dp}.
\end{align}}}

\subsection*{Operator $\widetilde{\mathcal{D}}_{2\alpha}$}
The spectral problem for this operator can be written as
{\small{\begin{align}
\left[ {\tau _3}{\partial _\sigma } - \frac{i}{2}\left( {\frac{1}{{\Omega\, {{\sinh }^2}\sigma }} + \Omega } \right){\tau _1}\right.  & - \frac{\alpha }{{2\, \Omega\, {{\cosh }^2}\left( {\sigma  + {\sigma _0}} \right)}}{\tau _2} \nonumber\\
&\left. - \frac{\alpha }{2}\left( {\tanh \left( {\sigma  + {\sigma _0}} \right) - \tanh \left( {2\sigma  + {\sigma _0}} \right)} \right)\mathbbm{1} \right]{\chi _{2\alpha }} =  \pm ip\, {\chi _{2\alpha }}.
\end{align}}}
The eigenfunctions satisfying this equation are given by a superposition of
\begin{align}
{Y_{2 \alpha }} &= {e^{ \pm i\alpha p\sigma }}\left( {{\delta _{\alpha , + }}\left[ {\begin{array}{*{20}{c}}
  {c_2^{\text{I}}} \\ 
  {c_2^{{\text{II}}}} 
\end{array}} \right] + {\delta _{\alpha , - }}\left[ {\begin{array}{*{20}{c}}
  {c_2^{{\text{II}}}} \\ 
  {c_2^{\text{I}}} 
\end{array}} \right]} \right),\\
{\bar Y_{2 \alpha }} &= {e^{ \mp i\alpha p\sigma }}\left( {{\delta _{\alpha , + }}\left[ {\begin{array}{*{20}{c}}
  {\bar{c}_2^{\text{I}}} \\ 
  {\bar{c}_2^{{\text{II}}}} 
\end{array}} \right] + {\delta _{\alpha , - }}\left[ {\begin{array}{*{20}{c}}
  {\bar{c}_2^{{\text{II}}}} \\ 
  {\bar{c}_2^{\text{I}}} 
\end{array}} \right]} \right),
\end{align}
where
\begin{align}
c_2^{\text I} &= \frac{{{2^{1/4}}\sqrt {\cosh \left( {\sigma  + {\sigma _0}} \right)} \left( { \pm i\alpha p - \frac{{\coth \sigma }}{2}} \right)}}{{{e^{{\sigma _0}{\text{/4}}}}{{\cosh }^{1/4}}\left( {2\sigma  + {\sigma _0}} \right)\left( { \pm i\alpha p - \frac{1}{2}} \right)}},\\
c_2^{\text{II}} &=  - \frac{{i\alpha {{\cosh }^{1/4}}{\sigma _0}}\sqrt {{\Omega\, \text{csch}\sigma}} }{{{2^{3/4}}\left( { \pm i\alpha p - \frac{1}{2}} \right){e^{{\sigma _0}{\text{/4}}}}}},\\
\bar c_2^{\text I} &=  - \frac{{i\alpha {e^{{\sigma _0}{\text{/4}}}}\cosh \left( {\sigma  + {\sigma _0}} \right)\sqrt {\Omega\, {\text{csch}}\sigma } }}{{{2^{5/4}}\left( { \pm i\alpha p{\text{ + }}\frac{1}{2}} \right){{\cosh }^{3/4}}{\sigma _0}\sqrt {\cosh \left( {2\sigma  + {\sigma _0}} \right)} }},\\
\bar c_2^{\text{II}} &= \frac{{{e^{{\sigma _0}{\text{/4}}}}{{\cosh }^{1/4}}\left( {2\sigma  + {\sigma _0}} \right)\left( { \pm i\alpha p{\text{ + }}\frac{{\coth \sigma }}{2}} \right)}}{{{2^{1/4}}\sqrt {\cosh \left( {\sigma  + {\sigma _0}} \right)} \left( { \pm i\alpha p{\text{ + }}\frac{1}{2}} \right)}}.
\end{align}
The asymptotic behaviour at large $\sigma$ of these solutions is given by
\begin{align}
\mathop {\lim }\limits_{\sigma  \to \infty } {Y_{2 \alpha }} &= {e^{ \pm i\alpha p\sigma }}\left( {{\delta _{\alpha , + }}\left[ {\begin{array}{*{20}{c}}
  1 \\ 
  0 
\end{array}} \right] + {\delta _{\alpha , - }}\left[ {\begin{array}{*{20}{c}}
  0 \\ 
  1 
\end{array}} \right]} \right),\\
\mathop {\lim }\limits_{\sigma  \to \infty } {{\bar Y}_{2 \alpha }} &= {e^{ \mp i\alpha p\sigma }}\left( {{\delta _{\alpha , + }}\left[ {\begin{array}{*{20}{c}}
  0 \\ 
  1 
\end{array}} \right] + {\delta _{\alpha , - }}\left[ {\begin{array}{*{20}{c}}
  1 \\ 
  0 
\end{array}} \right]} \right),
\end{align}
while for $\sigma\to0$ they have the following behaviour
\begin{align}
\mathop {\lim }\limits_{\sigma  \to 0} {Y_{2 \alpha }} &= \frac{{{v_2}}}{\sigma }\left( {{\delta _{\alpha , + }}\left[ {\begin{array}{*{20}{c}}
  1 \\ 
  {i\alpha } 
\end{array}} \right] + {\delta _{\alpha , - }}\left[ {\begin{array}{*{20}{c}}
  {i\alpha } \\ 
  1 
\end{array}} \right]} \right)+\mathcal{O}(\sigma^{0}),\\
\mathop {\lim }\limits_{\sigma  \to 0} {{\bar Y}_{2 \alpha }} &= \frac{{{{\bar v}_2}}}{\sigma }\left( {{\delta _{\alpha , + }}\left[ {\begin{array}{*{20}{c}}
  1 \\ 
  {i\alpha } 
\end{array}} \right] + {\delta _{\alpha , - }}\left[ {\begin{array}{*{20}{c}}
  {i\alpha } \\ 
  1 
\end{array}} \right]} \right)+\mathcal{O}(\sigma^{0}),
\end{align}
with
\begin{align}
{v_2} =  - \frac{{{{\left( {1 + \tanh {\sigma _0}} \right)}^{ - 1/4}}}}{{{2^{3/4}}\left( { \pm i\alpha p - \frac{1}{2}} \right)}},&&
{{\bar v}_2} =  - \frac{{i\alpha {{\left( {1 + \tanh {\sigma _0}} \right)}^{1/4}}}}{{{2^{5/4}}\left( { \pm i\alpha p + \frac{1}{2}} \right)}}.
\end{align}
A natural choice of boundary conditions is to pick the superposition which is well behaved at $\sigma\to0$. This fixes the superposition up to an overall constant
\begin{align}
{\chi _{2 \alpha }} \propto {Y_{2 \alpha }}{{\bar v}_2} - {{\bar Y}_{2 \alpha }}{v_2}.
\end{align}
Note that this superposition satisfies for both $\pm$ signs coming from $\omega=\pm i p$ and $\alpha$
\begin{align}
{\left. {{\Pi _ - }{\chi _{2\alpha }}} \right|_{\sigma  = 0}} = 0, && {\left. {{\Pi _ + }{\partial _\sigma }{\chi _{2\alpha }}} \right|_{\sigma  = 0}} = 0.
\end{align}
From the superposition, the phaseshift is given by
\begin{align}
{\delta _{2\, \alpha }} = \delta _{2\, \alpha }^ \pm =\pm\frac{\alpha}{2}{\rm{Arg}}\left(\frac{\bar{v}_{2}}{v_{2}} \right) = - \arctan 2p + \frac{\pi}{2} 
\end{align}
and the determinant is
\begin{align}
\ln {\det {{\widetilde{ \mathcal{D}}}_{2\alpha }}} &= \int\limits_0^\infty  {\tanh \pi p\left( {2\arctan 2p - \pi} -2Rp\right)dp}.
\end{align}

%

\subsection*{Operator $\widetilde{\mathcal{D}}_{3\alpha}$}
The spectral problem for this operator can be written as
{\small{
\begin{align}
\bigg[ {\tau _3}{\partial _\sigma } &+i \frac{{{\rm{sec}}{{\rm{h}}^2}\left( {\sigma  + {\sigma _0}} \right)}}{{2\Omega }}\left( {{\tau _1} + \alpha i{\tau _2}} \right)\nonumber\\
&\qquad\qquad - \frac{\alpha }{2}\left( {2 - \tanh \left( {\sigma  + {\sigma _0}} \right) - \tanh \left( {2\sigma  + {\sigma _0}} \right)} \right)\mathbbm{1} \bigg]{\chi _{3\alpha }} =  \pm ip\, {\chi _{3\alpha }}.
\end{align}}}
The eigenfunctions satisfying this equation are given by a superposition of
\begin{align}
{Y_{3 \alpha}} &= {e^{ \pm i\alpha p\sigma }}\left( {{\delta _{\alpha , + }}\left[ {\begin{array}{*{20}{c}}
  {c_3^{\text{I}}} \\ 
  0 
\end{array}} \right] + {\delta _{\alpha , - }}\left[ {\begin{array}{*{20}{c}}
  0 \\ 
  {c_3^{\text{I}}} 
\end{array}} \right]} \right),\\
{{\bar Y}_{3 \alpha }} &= {e^{ \mp i\alpha p\sigma }}\left( {{\delta _{\alpha , + }}\left[ {\begin{array}{*{20}{c}}
  {\bar c_3^{\text{I}}} \\ 
  {\bar c_3^{{\text{II}}}} 
\end{array}} \right] + {\delta _{\alpha , - }}\left[ {\begin{array}{*{20}{c}}
  {\bar c_3^{{\text{II}}}} \\ 
  {\bar c_3^{\text{I}}} 
\end{array}} \right]} \right),
\end{align}
where
\begin{align}
c_3^{\text{I}} &= \frac{{{e^\sigma }{e^{3{\sigma _0}/4}}}}{{{2^{3/4}}\sqrt {\cosh \left( {\sigma  + {\sigma _0}} \right)} {{\cosh }^{1/4}}\left( {2\sigma  + {\sigma _0}} \right)}},\\
\bar c_3^{\text I} &= \frac{{\alpha i\left( { \pm i\alpha p + 1 + \frac{{\coth \sigma }}{2}} \right){e^{ - \sigma  - 3{\sigma _0}/4}}{\Omega ^{1/2}}{\text{sec}}{{\text{h}}^{3/4}}{\sigma _0}{{\sinh }^{3/2}}\sigma }}{{\left( { \pm i\alpha p{\text{ + }}\frac{3}{2}} \right)\left( { \pm i\alpha p{\text{ + }}\frac{1}{2}} \right){2^{1/4}}\sqrt {\cosh \left( {2\sigma  + {\sigma _0}} \right)} }},\\
\bar c_3^{{\text{II}}} &= \frac{{{2^{3/4}}{e^{ - \sigma  - 3{\sigma _0}/4}}\sqrt {\cosh \left( {\sigma  + {\sigma _0}} \right)} }}{{{\text{sec}}{{\text{h}}^{1/4}}\left( {2\sigma  + {\sigma _0}} \right)}}.
\end{align}
The asymptotic behaviour at large $\sigma$ of these solutions is given by
\begin{align}
\mathop {\lim }\limits_{\sigma  \to \infty } {Y_{3 \alpha }} &= {e^{ \pm i\alpha p\sigma }}\left( {{\delta _{\alpha , + }}\left[ {\begin{array}{*{20}{c}}
  1\\ 
  0 
\end{array}} \right] + {\delta _{\alpha , - }}\left[ {\begin{array}{*{20}{c}}
  0 \\ 
  1 
\end{array}} \right]} \right),\\
\mathop {\lim }\limits_{\sigma  \to \infty } {{\bar Y}_{3 \alpha }} &= {e^{ \mp i\alpha p\sigma }}\left( {{\delta _{\alpha , + }}\left[ {\begin{array}{*{20}{c}}
  0 \\ 
  1 
\end{array}} \right] + {\delta _{\alpha , - }}\left[ {\begin{array}{*{20}{c}}
  1 \\ 
  0 
\end{array}} \right]} \right),
\end{align}
while for $\sigma\to0$ they have the following behaviour
\begin{align}
\mathop {\lim }\limits_{\sigma  \to 0} {Y_{3 \alpha }} &= {v_3}\left( {{\delta _{\alpha , + }}\left[ {\begin{array}{*{20}{c}}
1\\
0
\end{array}} \right] + {\delta _{\alpha , - }}\left[ {\begin{array}{*{20}{c}}
0\\
1
\end{array}} \right]} \right) + \mathcal{O}\left( \sigma  \right),\\
\mathop {\lim }\limits_{\sigma  \to 0} {\bar Y_{3 \alpha }} &= {v_{3}^{-1}}\left( {\delta _{\alpha , + }}\left[ {\begin{array}{*{20}{c}}
{i\, \frac{\alpha }{4}\,  {\text{sech}^2}{\sigma _0}\ {{\left( { \pm i\alpha p + \frac{1}{2}} \right)}^{ - 1}}{{\left( { \pm i\alpha p + \frac{3}{2}} \right)}^{ - 1}}}\\
1
\end{array}} \right] \right.\nonumber\\
& \left. \qquad\quad + {\delta _{\alpha , - }}\left[ {\begin{array}{*{20}{c}}
1\\
{i\, \frac{\alpha }{4}\, \text{sech}^2 {\sigma _0}\ {{\left( { \pm i\alpha p + \frac{1}{2}} \right)}^{ - 1}}{{\left( { \pm i\alpha p + \frac{3}{2}} \right)}^{ - 1}}}\end{array}} \right] \right)+\mathcal{O}\left(\sigma\right),
\end{align}
with
\begin{align}
{v_3} = {\left( {\frac{{1 + \tanh {\sigma _0}}}{2}} \right)^{3/4}} .
\end{align}
Imposing the boundary conditions
\begin{align}
{\left. {{\Pi _ - }{\chi _{3\alpha }}} \right|_{\sigma  = 0}} = 0, && {\left. {{\Pi _ + }{\partial _\sigma }{\chi _{3\alpha }}} \right|_{\sigma  = 0}} = 0,
\end{align}
fixes the superposition (up to an overall constant) to be
\begin{align}
{\chi _{3 \alpha }} \propto {Y_{3 \alpha }}{{\bar v}_3} - {{\bar Y}_{3 \alpha }}{v_3},
\end{align}
where
{\small{\begin{align}
{\bar v_3} =  - \frac{i\alpha\, \text{sech}^2\sigma _0\, {{\left( {\frac{{1 + \tanh {\sigma _0}}}{2}} \right)}^{ - 3/4}}}{{2\left( {{p^2} + \frac{1}{4}} \right)\left( {{p^2} + \frac{9}{4}} \right)}}\left[ { \pm \alpha ip - 2{{\cosh }^2}{\sigma _0}\left( {{p^2} + \frac{1}{4}} \right)\left( {{p^2} + \frac{9}{4}} \right) + \frac{1}{2}\left( {{p^2} - \frac{3}{4}} \right)} \right].
\end{align}}}
From the superposition, the phaseshift is given by
\begin{align}
{\delta _{3\, \alpha }} = \delta _{3\, \alpha }^ \pm= \pm \frac{\alpha }{2}\text{Arg}\left( {\frac{{{{\bar v}_3}}}{{{v_3}}}} \right)  = \frac{1}{2}\arctan \frac{p}{{ - 2{{\cosh }^2}{\sigma _0}\left( {{p^2} + \frac{1}{4}} \right)\left( {{p^2} + \frac{9}{4}} \right) + \frac{1}{2}\left( {{p^2} - \frac{3}{4}} \right)}}
\end{align}
and the determinant is
{\footnotesize{\begin{align}
\ln {\det {{\widetilde{\mathcal D}}_{3\alpha }}} \mathop  = \int\limits_0^\infty  {\tanh \pi p\left( -\arctan \frac{p}{{ - 2{{\cosh }^2}{\sigma _0}\left( {{p^2} + \frac{1}{4}} \right)\left( {{p^2} + \frac{9}{4}} \right) + \frac{1}{2}\left( {{p^2} - \frac{3}{4}} \right)}} -2Rp \right)dp} .
\end{align}}}

\section{Collecting all the pieces}\label{sec5}
The semiclassical string partition function results from collecting all the contributions entering \eqref{TheZ}, obtaining
{\small{\begin{align}\label{clnZ}
\ln Z \left( {{\sigma _0}} \right) &= \int\limits_0^\infty  dp\Bigg[ \coth \pi p\ \Bigg( \arctan \frac{p}{{1 + \tanh {\sigma _0}}} + 2\arctan \frac{{2p}}{{1 + \tanh {\sigma _0}}} - \arctan \frac{p}{2} \nonumber\\
&\left.\,  - 4\arctan p + \pi  \Bigg) - \tanh \pi p\ \Bigg( 2\arctan \frac{p}{{\frac{1}{2} + \tanh {\sigma _0}}} - 6\arctan 2p\right.\nonumber\\ 
&\left.\, - 2\arctan \frac{{2p}}{3} + \arctan \frac{p}{{ - 2{{\cosh }^2}{\sigma _0}\left( {{p^2} + \frac{1}{4}} \right)\left( {{p^2} + \frac{9}{4}} \right) + \frac{1}{2}\left( {{p^2} - \frac{3}{4}} \right)}}   + 3\pi  \Bigg)\right.\nonumber\\
&\, + 8Rp\left( {\coth \pi p - \tanh \pi p} \right) \Bigg] .
\end{align}}}
Using equation \eqref{clnZ}, the ratios of 1-loop string partitions functions can be evaluated by direct integration. The easiest way to do this is by first differentiating with respect to $\sigma_0$
{\small{\begin{align}
\frac{d}{{d{\sigma _0}}}\ln Z\left( {{\sigma _0}} \right) &= \frac{1}{{{{\cosh }^2}{\sigma _0}}}\int\limits_0^\infty  {dp}\, p\left[ 2\frac{{\tanh \pi p}}{{{p^2} + {{\left( {\frac{1}{2} + \tanh {\sigma _0}} \right)}^2}}} + \frac{1}{2}\frac{{\tanh \pi p}}{{{p^2} + {{\left( {1 + \frac{{\tanh {\sigma _0}}}{2}} \right)}^2}}}\right.\nonumber\\
&\left.\ \  - \frac{1}{2}\frac{{\tanh \pi p}}{{{p^2} + {{\left( {1 - \frac{{\tanh {\sigma _0}}}{2}} \right)}^2}}} - \frac{{\coth \pi p}}{{{p^2} + {{\left( {1 + \tanh {\sigma _0}} \right)}^2}}} - \frac{{\coth \pi p}}{{{p^2} + {{\left( {\frac{{1 + \tanh {\sigma _0}}}{2}} \right)}^2}}} \right]\nonumber\\
&\ \ + \frac{{dR}}{{d{\sigma _0}}}.
\end{align}}}
Using the following identities
\begin{align}
\tanh \pi p = 1 - \frac{2}{{{e^{2\pi p}} + 1}}, \qquad\quad &\qquad\quad \coth \pi p = 1 + \frac{2}{{{e^{2\pi p}} - 1}},\\
\int\limits_0^\infty  \frac{{dp\, p}}{{\left( {{e^{2\pi p}} + 1} \right)\left( {{p^2} + {c^2}} \right)}}& =  - \frac{{\ln c}}{2} + \frac{1}{2}\psi  \left( {c + \frac{1}{2}} \right),\\
\int\limits_0^\infty  \frac{{dp\, p}}{{\left( {{e^{2\pi p}} - 1} \right)\left( {{p^2} + {c^2}} \right)}} &= \frac{{\ln c}}{2} - \frac{1}{{4c}} - \frac{1}{2}\psi \left( c \right),
\end{align}
the integral over $p$ reduces to
{\small{\begin{align}
\frac{d}{{d{\sigma _0}}}\ln Z\left( {{\sigma _0}} \right) = \frac{{\text{sech}^{2}{\sigma _0}}}{2}\left( {{H_{\frac{{1 - \tanh {\sigma _{0}}}}{2}}} + {H_{\frac{{1 + \tanh {\sigma _{0}}}}{2}}} - 2{H_{1 + \tanh {\sigma _{0}}}}} \right) +\frac{1-\tanh\sigma_{0}}{2}+ \frac{{dR}}{{d{\sigma _{0}}}}.
\end{align}}
Replacing the diffeomorphism invariant regulator \eqref{IRcutoff}, the last two terms on the right hand side cancel and integration over $\sigma_{0}$ results in
\begin{align}
\ln \frac{{Z\left( \infty  \right)}}{{Z\left( {{\sigma _0}} \right)}} = \ln \Gamma \left( {2 + \tanh {\sigma _0}} \right) + \ln \Gamma \left( {\frac{{3 - \tanh {\sigma _0}}}{2}} \right) - \ln \Gamma \left( {\frac{{3 + \tanh {\sigma _0}}}{2}} \right) - \ln 2.
\end{align}
In terms of the angle $\theta_{0}$, the semiclassical contribution to the ratio is\footnote{This result can be rewritten in a form closer to the expressions presented in \cite{Aguilera-Damia:2018bam}
$$\ln \frac{{Z\left( 0 \right)}}{{Z\left( {{\theta _0}} \right)}} =  - \ln \left( {\frac{{4\cos \left( {\frac{{\pi \cos {\theta _0}}}{2}} \right)}}{{\pi\, {{\sin }^2}{\theta _0}}}} \right) - 2\ln \cos \frac{{{\theta _0}}}{2} - 2\ln \Gamma \left( {{{\cos }^2}\frac{{{\theta _0}}}{2}} \right) + \ln \Gamma \left( {1 + \cos {\theta _0}} \right),$$
where the first term on the right comes from the massless fermionic operator $\widetilde{\mathcal{D}}_{3\alpha}$.}
\begin{align}\label{semipiece}
\ln \frac{{Z\left( 0 \right)}}{{Z\left( {{\theta _0}} \right)}} = \ln \Gamma \left( {2 + \cos {\theta _0}} \right) + \ln \Gamma \left( {\frac{{3 - \cos {\theta _0}}}{2}} \right) - \ln \Gamma \left( {\frac{{3 + \cos {\theta _0}}}{2}} \right) - \ln 2.
\end{align}
Collecting the classical and semiclassical contributions \eqref{classypiece} and \eqref{semipiece}, respectively, leads to
\begin{align}\label{st1loop}
\frac{{\left\langle {W_F^{\frac{1}{2}}\left( 0 \right)} \right\rangle }}{{\left\langle {W_F^{\frac{1}{6}}\left( {{\theta _0}} \right)} \right\rangle }} =\frac{{{Z_{{\rm{string}}}}\left( 0 \right)}}{{{Z_{{\rm{string}}}}\left( {{\theta _0}} \right)}} = {e^{\pi \sqrt {2\lambda } \left( {1 - \cos {\theta _0}} \right)}}\ \frac{{\Gamma \left( {2 + \cos {\theta _0}} \right)\Gamma \left( {\frac{{3 - \cos {\theta _0}}}{2}} \right)}}{{2\Gamma \left( {\frac{{3 + \cos {\theta _0}}}{2}} \right)}},
\end{align}
in perfect agreement with the field theory result \eqref{Theprediction}.

\section{The $\frac{1}{2}$-BPS fermionic circular Wilson loop}\label{sec6}

We now consider the string theory computation in $AdS_{4}\times CP^{3}$ of the $\frac{1}{2}$-BPS fermionic circular Wilson loop in ABJM at next to leading order at strong coupling. To do this, we consider the ratio between this Wilson loop with $\theta_{0}=0$ and the $\frac{1}{6}$-BPS latitude Wilson loop with angle $\theta_{0}=\frac{\pi}{2}$ and trivial expectation value (recall equation \eqref{wtrivial}).

For the case of $AdS_{5}\times S^{5}$, the analogous computation in \cite{Medina-Rincon:2018wjs} lead to successful matching with the field theory prediction for the $\frac{1}{2}$-BPS circle ($\theta_{0}=0$) in $\mathcal{N}=4$ SYM: {\footnotesize{$\ln \left\langle W_{\rm{C}} \right\rangle  = \sqrt \lambda   - \frac{3}{4}\ln \lambda  + \frac{1}{2}\ln \frac{2}{\pi } + \mathcal{O}\left( {{\lambda ^{ - 1/2}}} \right)$}}. In string theory the $\sqrt{\lambda}$ term comes from the regularized minimal area, the $\ln\lambda$ is commonly attributed to three ghost zero modes, while the $\lambda^{0}$ term comes from ghost/longitudinal modes and string fluctuations. Meanwhile, for the ($\theta_{0}=\frac{\pi}{2}$) $\frac{1}{4}$-BPS ``special" latitude Wilson loop one has that $\ln \left\langle W_{\rm{L}} \right\rangle =0$, where the $\sqrt{\lambda}$ term is zero due to the areas over $AdS$ and $S$ canceling each other, while cancellation of the $\ln\lambda$ term comes from the three ghost zero modes being cancelled by three bosonic zero modes due to degeneracies of the classical solution at $\theta_{0}=\frac{\pi}{2}$ \cite{Zarembo:2002an}.

When doing the $AdS_{5}\times S^{5}$ computation of \cite{Medina-Rincon:2018wjs}, ghost and longitudinal contributions are assumed to cancel in the ratio and thus are not taken into account. Consequently, the ratio of Wilson loops with angles $\theta_{0}=0$ and $\theta_{0}=\frac{\pi}{2}$ results from both considering the three bosonic zero modes from the moduli of the classical solution at $\theta_{0}=\frac{\pi}{2}$ and from string fluctuations given in terms of phaseshifts and regulators. Just considering only the contributions to the determinants in terms of phaseshifts and IR regulators leads to a divergent result for the ratio of the $\theta_{0}=0$ and $\theta_{0}=\frac{\pi}{2}$ Wilson loops\footnote{It is important to note that in $AdS_{5}\times S^{5}$ the phaseshifts and IR regulators fully account for the ratio between the $\theta_{0}=0$ and $\theta_{0}\in\left[0,\frac{\pi}{2}\right)$ latitudes leading to a finite result \cite{Cagnazzo:2017sny}. In such calculation the relative number of zero modes between the Wilson loops is null since the classical solution is only degenerate for $\theta_{0}=\frac{\pi}{2}$.}. This divergence is compensated by the zero modes of the moduli leading to a finite end result \cite{Medina-Rincon:2018wjs}. Effectively, the moduli  account for the $\ln\lambda$ term in the logarithm of the ratio of Wilson loops, while the $\lambda^{0}$ term results after careful evaluation of both the moduli and fluctuation contributions.

For the case of fermionic latitude Wilson loops in $AdS_{4}\times CP^{3}$ the situation is quite different. From the field theory predictions \eqref{wlatitude}, \eqref{wcircle} and \eqref{wtrivial}, we see that there are no $\ln\lambda$ terms appearing in the logarithm of the ratio of any two latitude Wilson loops with angles $\theta_{0}\in\left[0,\frac{\pi}{2}\right]$. This suggests that in string theory at 1-loop the relative number of zero modes between any two fermionic latitudes is zero\footnote{Furthermore, inspection of the classical solution in $AdS_{4}\times CP^{3}$ at $\theta_{0}=\frac{\pi}{2}$ does not seem to lead to bosonic zero modes, unlike in the $AdS_{5}\times S^{5}$ case. From the perspective of the bosonic differential operators, $\mathcal{K}_{1}$ and $\mathcal{K}_{3\alpha}$ also appeared in the $AdS_{5}\times S^{5}$ case where they had no zero mode contributions, while $\mathcal{K}_{2\alpha}$ is easily related to $\mathcal{K}_{3\alpha}$ after doing $\sigma\to2\sigma$ and $\tau\to2\tau$.}. 

In this paper we have assumed cancellation between ghost/longitudinal mode contributions when considering ratios of Wilson loops in $AdS_{4}\times CP^{3}$. Assuming that additional zero modes (if any) appearing for individual Wilson loops are cancelled when considering ratios, the entire answer is given only in terms of phaseshifts and diffeomorphism invariant regulators. Explicit integration of the ratio of non-zero mode semiclassical contributions for $\theta_{0}=0$ and $\theta_{0}=\frac{\pi}{2}$ coming from \eqref{clnZ}, unlike in the $AdS_{5}\times S^{5}$ case, indeed results in a finite answer. Direct evaluation of the required expressions for $\theta_{0}=\frac{\pi}{2}$ ($\sigma_{0}=0$), or equivalently taking the limit $\theta_{0}\to\frac{\pi}{2}$ in \eqref{st1loop}, leads to
\begin{align}
\left\langle {W_F^{\frac{1}{2}}} \right\rangle  = \frac{{{Z_{{\rm{string}}}}\left( 0 \right)}}{{{Z_{{\rm{string}}}}\left( {\frac{\pi }{2}} \right)}} = \frac{1}{2}{e^{\pi \sqrt {2\lambda } }},
\end{align}
which matches the localization result for the $\frac{1}{2}$-BPS fermionic circular Wilson loop in ABJM \cite{Drukker:2010nc}.


\section{Conclusions}\label{sec7}
In this paper we have shown how quantum string corrections in $AdS_{4}\times CP^{3}$ reproduce the expectation value at next to leading order at strong coupling in ABJM for the ratio of $\frac{1}{6}$-BPS latitude and $\frac{1}{2}$-BPS circular fermionic Wilson loops, as well as the known localization result for the $\frac{1}{2}$-BPS circular Wilson loop. 

For the $\frac{1}{2}$-BPS circular Wilson loop we considered its ratio with the latitude Wilson loop with trivial expectation value and assumed that zero mode contributions  of individual Wilson loops cancel in the ratio. In string theory it is unclear how this latitude Wilson loop has trivial expectation value beyond leading order as the usual counting argument for C.K.V.'s and moduli has to be revised. Despite there being a qualitative argument for the analogous configuration in $AdS_{5}\times S^{5}$ \cite{Zarembo:2002an}, precise derivations for these individual loops are missing in string theory.

Another open question concerning Wilson loops in $AdS_{4}\times CP^{3}$ is the matching with field theory results for Wilson loops with winding \cite{Klemm:2012ii,Okuyama:2016deu,ccOkuyama}. It would be compelling to understand such winding contributions in string theory, although attempts for the $AdS_{5}\times S^{5}$ case have yet to agree with field theory \cite{Kruczenski:2008zk,Bergamin:2015vxa}. Finally, natural steps towards a better understanding of Wilson loops in string theory would be to compute individual circular Wilson loops instead of ratios and to extend current techniques beyond 1-loop.

\appendix


\subsection*{Acknowledgements}
The author would like to thank J. Aguilera-Damia, M. Bianchi, L. Griguolo, B. Hoare, M. Mari\~no, K. Okuyama, D. Seminara, A. Tseytlin, L. Wulff and K.~Zarembo for interesting discussions and/or correspondence. The author would also like to thank Nordita (the Nordic Institute for Theoretical Physics) for generous hospitality during parts of this work. This work was partially supported by grant no. 615203 from the European Research Council under the FP7 and by the Swiss National Science Foundation through the NCCR SwissMAP.


\bibliographystyle{nb}
\bibliography{refs}

\end{document}